# Imaging superconducting weak spots through vortex-assisted THz near-field photovoltage

S.J. Salvía-Fernández[1], E. Khestanova[1], Z. Velluire-Pellat[1], L.R. Cadorim[2,3], S. Battle-Porro[1], D.A. Czaplewski[4], J. Osmond[1], S. Castilla[1], A. Bachtold[1], M. V. Milošević[2,3], A. Kapitulnik[5,6,7,8], F.H.L. Koppens[1,9]

1. ICFO – Institut de Ciències Fotòniques, The Barcelona Institute of Science and Technology, Castelldefels (Barcelona), Spain.
2. COMMIT, Department of Physics, University of Antwerp, Groenenborgerlaan 171, B-2020, Antwerp, Belgium.
3. Stavropoulos Center for Complex Quantum Matter, Department of Physics and Astronomy, University of Notre Dame, Notre Dame, IN, USA.
4. Center for Nanoscale Materials, Argonne National Laboratory, Lemont, IL 60439, USA.
5. Geballe Laboratory for Advanced Materials, Stanford University, Stanford, CA, USA.
6. Stanford Institute for Materials and Energy Sciences, SLAC National Accelerator Laboratory, Menlo Park, CA, USA.
7. Department of Applied Physics, Stanford University, Stanford, CA, USA.
8. Department of Physics, Stanford University, Stanford, CA, USA.
9. ICREA – Institució Catalana de Recerca i Estudis Avançats, Barcelona, Spain.

## Abstract

Nanoscale inhomogeneities are a defining feature of many superconducting materials, yet their local electromagnetic response has remained difficult to access experimentally. This is because their relevant energy scale lies in the terahertz range, where wavelengths — on the order of hundreds of microns — are too large to spatially resolve nanoscopic variations. Here, we demonstrate the first application of THz near-field photovoltage nanoscopy in a superconductor, achieving 300 nm spatial resolution at 2.52 THz. Scanning a current-biased NbN strip, we reveal photovoltage peaks within the bulk associated with nanoscopic defects of reduced superfluid density. The observed photovoltage follows the evolution of the vortex-dissipative state and is attributed to enhanced vortex–antivortex pair nucleation at defect sites. Together, these results open a direct route to probing how material inhomogeneities influence light–matter interactions in superconductors, with implications for superconducting devices and strongly inhomogeneous systems such as high-Tc and moiré materials.

# Introduction

Nanoscale disorder often determines where a superconductor first becomes dissipative. In thin film superconductors, compositional variations, grain boundaries and stoichiometric defects lead to local regions where the order parameter is suppressed and where vortices are more readily nucleated under current bias. Such weak superconducting spots directly affect device operation, from the position-dependent detection efficiency of superconducting nanowire single-photon detectors[1,2] to defect-limited coherence in transmon qubits[3]. In strongly inhomogeneous quantum materials, spatial variations of superconductivity are not only extrinsic imperfections but part of the electronic ground state itself, as in electronic phase separation in high-Tc superconductors[4] and correlated moiré systems[5]. Therefore, the central challenge is to probe superconducting inhomogeneity in the regime where it matters most, that is, at superconductivity-relevant energy scales, on sub-micron length scales, and under the current-biased conditions in which weak spots generate dissipation.

The relevant energy scale of the problem is in the range of millielectronvolts (meV), that is sub-terahertz and terahertz electromagnetic frequencies. Superconducting gaps commonly lie in the microwave-THz range[6,7]and THz spectroscopy has become a key probe of superconducting electrodynamics, including collective modes[8–11], gap symmetry[6,7,12] and non-equilibrium superconducting states[13]. However, THz wavelengths are macroscopic compared with the disorder landscape: at 3 THz, the diffraction-limited spatial resolution is ∼100 μm, whereas defects, grains, weak links and phase-separated regions can be 10-1000 nm in size. This energy-lengthscale mismatch has prevented direct access to the local THz response of individual superconducting weak spots. Resolving this problem requires a probe that combines THz-frequency excitation with nanoscale spatial confinement and an electrical readout sensitive to the onset of dissipation.

Here we address this mismatch using THz near-field photovoltage nanoscopy[14] in a current-biased superconductor. The used metallic AFM tip confines 2.52 THz radiation to a nanoscale hotspot, achieving 300 nm spatial resolution, almost 400 times below the far-field diffraction limit. In NbN, this THz photon energy exceeds the superconducting gap, so the near field acts as a local pair-breaking perturbation. The applied DC current then converts the local suppression of superconductivity into a measurable photovoltage through vortex-assisted dissipation. Scanning across an NbN strip, we observe reproducible bulk photovoltage peaks fixed in sample coordinates. We interpret these peaks as vortex-nucleation weak spots, in particular, defect-associated regions of reduced condensate robustness where the THz hotspot preferentially enhances vortex-antivortex pair creation. Their non-monotonic bias-current dependence tracks the vortex-dissipative state, and time-dependent Ginzburg-Landau simulations support a mechanism in which local order-parameter suppression enhances vortex entry at the edges and vortex-antivortex nucleation at the defects. These results establish THz near-field photovoltage nanoscopy as a functional probe of superconducting weak spots under current bias, directly linking nanoscale inhomogeneity to light-driven vortex dissipation in superconducting films and devices.

# Results

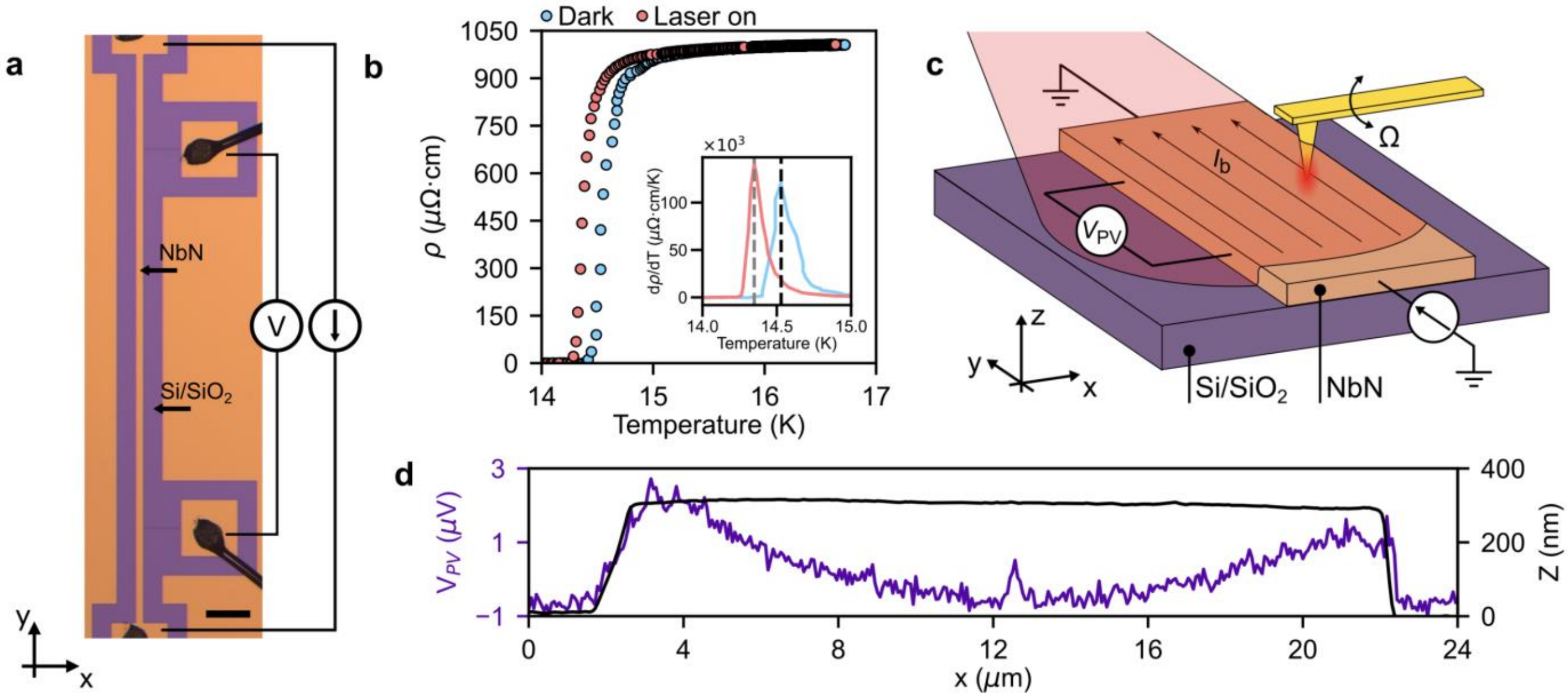

**Fig. 1. Electron transport and near-field photovoltage profiling under THz illumination. a** Microscope image of the device (scale bar 150 µm). Orange regions correspond to the NbN film, purple corresponds to the $Si/SiO_2$ substrate. **b** Device resistivity, $\rho$, as a function of temperature measured in the dark and under continuous-wave 2.52 THz illumination with the laser power of 1.7 mW at the cryostat window. The inset shows the temperature derivative d$\rho$/dT, with dashed lines marking the superconducting transition temperature ($T_c$) defined at the maximum of the derivative. **c** Schematic diagram of the THz near-field photovoltage experiment. **d** THz near-field photovoltage (purple line) and AFM topography (black line) profiles obtained at $T = 13.86$ K by averaging in the DC current range $I_\mathrm{b} = [380, 420]$ µA.

## Sample characterization

Fig. 1a shows an optical micrograph of the device under study. A NbN film with thickness $d = 320$ nm was sputter-deposited on a $Si/SiO_2$ substrate. The scanning electron microscopy (SEM) images (Supplementary Note 1) reveal a granular morphology, with triangular grains of typical size $\sim$ 15 - 35 nm, characteristic of (111)-orientated NbN films[15], as confirmed by X-ray diffraction (Fig. S2). The final structure consists of a strip with width $w = 20$ µm terminated by two contact pads for current sourcing and two voltage probes, separated by a length $L = 1$ mm. In the normal state, the sample resistivity $\rho_\mathrm{n}(17\ \mathrm{K}) = 1006\ \mu\Omega\cdot\mathrm{cm}$ (Fig. 1b) is close to the values found in granular NbN films[15]. The film exhibits a sharp superconducting transition at $T_\mathrm{c}^\mathrm{Dark} = 14.53$ K (defined as the $d\rho/dT$ maximum), with width $\Delta T^\mathrm{Dark} = 0.3 \pm 0.03$ K (defined between 90% and 10% of $\rho_n(17\ \mathrm{K})$).

A 2.52 THz laser beam was collimated to deliver 1.7 mW of power at the cryostat window and focused onto the sample using a parabolic mirror. Under far-field THz exposure, the critical temperature is lowered to $T_\mathrm{c}^{2.52\mathrm{THz}} =$ 14.35 K, as shown in Fig. 1b. Since the photon energy ($\hbar\omega = 10.4$ meV) exceeds the NbN superconducting gap ($2\Delta \approx 5.8$ meV)[16], photon absorption induces Cooper pair breaking and quasiparticle excitation, increasing the electronic temperature. This bolometric response is observed as a reduction of the critical temperature of $\Delta T_\mathrm{c} \approx$ 0.18 K, that leaves the width of the superconducting transition ($\Delta T^{2.52\mathrm{THz}} = 0.27 \pm 0.03$ K) unchanged.

## THz Near-field photovoltage nanoscopy

To probe the local THz response of the device shown in Fig. 1a with nanometric resolution we employ THz near-field photovoltage nanoscopy (Fig. 1c). A gold-coated AFM tip operated in tapping mode under THz illumination acts as a nanoantenna[17–19], confining the incident 2.52 THz radiation to a volume set by the tip radius ($\sim$ 250 nm). This yields a spatial resolution of $\sim$100 - 300 nm, an improvement of $\sim$ 300 – 1000 times over the far-field diffraction limit ($\sim$ 100 µm at this frequency). The measured quantity (the near-field photovoltage, $V_\mathrm{PV}$) is the change in the global voltage across the strip, detected at the tapping frequency Ω[20–22], that results from the local near-field perturbation while a DC bias current, $I_\mathrm{b}$, is applied.

Fig. 1d shows a $V_{\mathrm{PV}}$ profile measured at $T = 13.86$ K under DC current bias. The profile was obtained by averaging near-field line scans over the current range $I_{\mathrm{b}} = [380, 420]$ µA. The DC current was applied along the y-axis while the tip was scanned perpendicular to the current flow, across the width of the strip. The simultaneously acquired AFM topography image exhibits a flat central region and a step of 320 nm, consistent with the device cross section. In contrast, the $V_{\mathrm{PV}}$ trace shows a pronounced spatial dependence, featuring an enhanced response near the edges and a localized peak in the central region, at $x = 12.52$ µm.

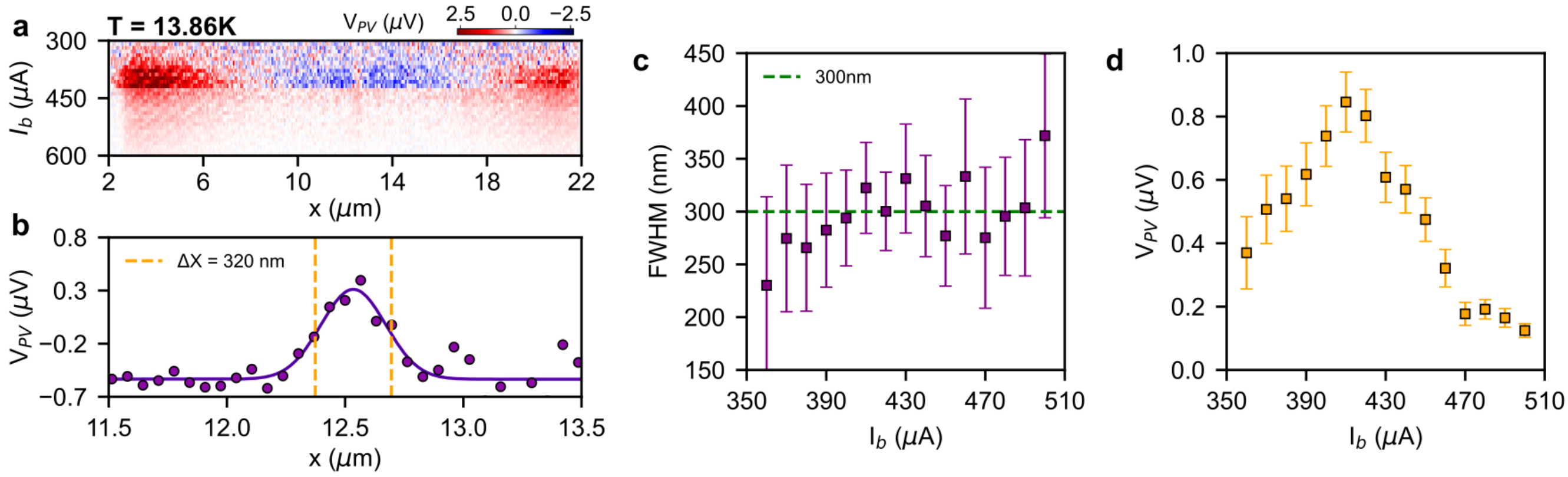

**Fig. 2. Local peak in the near-field photovoltage at $T = 13.86$ K. a** Current-space near-field photovoltage map measured at $T = 13.86$ K. **b** Zoomed-in view of the $V_{\mathrm{PV}}$ profile obtained by averaging the line scans from the map in (a) over the current range $I_{\mathrm{b}} = [380, 420]$ µA, with a Gaussian fit to the photovoltage peak (solid purple line) found at $x = 12.52$ µm. Orange vertical dashed lines indicate the FWHM of the peak. **c** Current dependence of the FWHM of the peak extracted from the map in (a). **d** Current dependence of amplitude of the central peak extracted from the analysis of the map in panel (a).

We first focus on the local $V_{\mathrm{PV}}$ peak within the bulk. To study it further, we recorded $V_{\mathrm{PV}}$ profiles at different DC currents, $I_{\mathrm{b}}$. The traces are combined to form a current-space near-field photovoltage map, $V_{\mathrm{PV}}(x, I_{\mathrm{b}})$ (Fig. 2a). The obtained $V_{\mathrm{PV}}$ map shows that the local photovoltage response is present for a broad range of bias currents, appearing at a fixed position ($x = 12.52$ µm). As shown in the zoomed-in view of the $V_{\mathrm{PV}}$ trace (Fig. 2b) this feature rises above the noise level and can be fitted using a Gaussian profile from which we extract the FWHM and the amplitude of the local signal as a function of $I_{\mathrm{b}}$. The found FWHM is approximately constant with current, with an average FWHM $\langle \Delta x \rangle = 300 \pm 30$ nm (Fig. 2c). The amplitude of the peak follows a pronounced non-monotonic trend (Fig. 2d), reaching a maximum at $I_{\mathrm{b}} = 410$ µA.

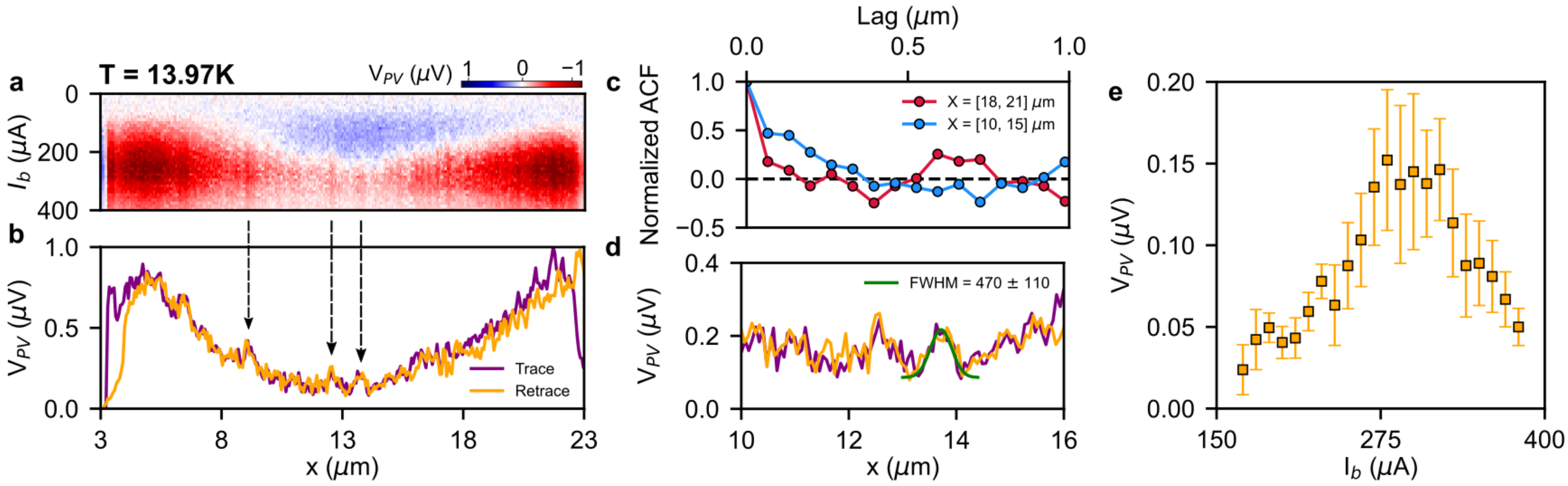

**Fig. 3. Multiple local near-field photovoltage peaks at $T = 13.97$ K. a** Current-space near-field photovoltage map measured at $T = 13.97$ K (full map shown in Fig. S6). **b** Horizontal $V_{\mathrm{PV}}$ profiles (trace and retrace) obtained by averaging the line scans from the map in panel (a) over the current range $I_{\mathrm{b}} = [290, 350]$ µA. The trace and retrace profiles were shifted by 1 µm to account for hysteresis in the piezoelectric scanners. **c** Zoomed-in view of the trace and retrace $V_{\mathrm{PV}}$ profiles shown in (b). The green line plots the Gaussian fit to the peak in the trace profile. **d** Normalized autocorrelation

functions computed for two different regions taken from the PV trace signal shown in (b). **e** Current dependence of the amplitude of the local peak fitted in (c).

The presence of local peaks of $V_{\mathrm{PV}}$ in the bulk of the strip is confirmed by measuring a current-space map at a different sample location and temperature, shown in Fig. 3a. The $V_{\mathrm{PV}}$ map reveals again several regions of increased photovoltage, appearing as vertical lines in the current-space maps. These features are reproducible between trace (increasing $x$) and retrace (decreasing $x$) scans, as demonstrated by the clear overlap of the local peaks (Fig. 3b., dashed arrows). To ensure that the observed features are genuine localized peaks and not noise, we compute the autocorrelation function (ACF) of the $V_{\mathrm{PV}}$ trace in different regions (Fig. 3c, more details in Supplementary Note 5). The central region ($x$ = [10, 15] µm), which contains two peaks highlighted in Fig. 3b, exhibits a slowly decaying ACF as expected for spatially correlated features. In contrast, in an off-central region ($x$ = [18, 21] µm) the ACF decays to the noise level after the first lag, indicating that the observed spikes in $V_{\mathrm{PV}}$ in this region are related to noise, as confirmed by the absence of overlap between trace and retrace. We fit the signal of one representative $V_{\mathrm{PV}}$ peak again using a Gaussian (Fig. 3d), yielding a FWHM of 470 ± 110 nm, broader than the peak obtained at 13.86 K (300 ± 30 nm). Similar to what was observed in Fig. 2d, the current evolution of the peak follows a non-monotonic dependence on current (Fig. 3e), reaching a maximum around $I_{\mathrm{b}} = 300$ µA.

The reproducible localized peaks in $V_{\mathrm{PV}}$ observed across multiple locations, temperatures, and bias currents, presented in Fig. 2 and Fig. 3 demonstrate the presence of regions within the bulk of the superconducting strip that interact with the THz near-field generating a photovoltage signal. We ascribe this response to nanometric defects within the NbN strip with locally weakened superconductivity. Possible origins for these defects will be further addressed in the discussion section.

## Vortex-assisted near-field photovoltage mechanism

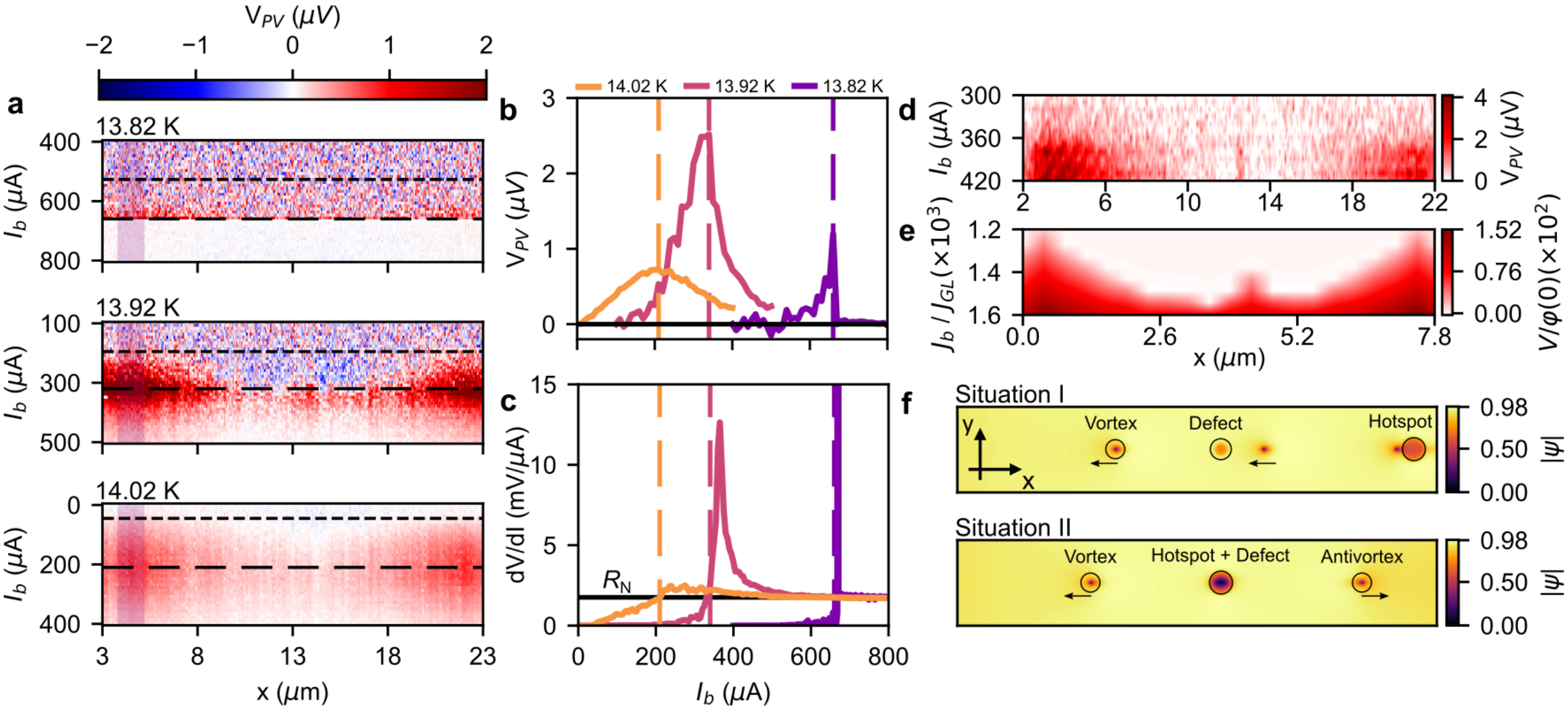

**Fig. 4. Current dependence of the near-field photovoltage signal. a** Current-space $V_{\mathrm{PV}}$ maps measured at various temperatures. Short-dashed lines mark the $I_c$ values obtained from the far-field VI curves using the criterion $V > 50$ µV. Long-dashed lines indicate the current corresponding to the maximum observed photovoltage. The purple shaded area corresponds to the region used to extract the horizontal cuts shown in (b). **b** Near-field photovoltage at the edge as a function of bias current, extracted from the maps in panel (a). Vertical dashed lines indicate the current of maximum $V_{\mathrm{PV}}$. **c** dV/dI curves extracted from the DC transport data under far-field illumination. **d** Zoomed-in current-space map taken at T = 13.86K showing a local photovoltage peak close to the center of the microstrip. **e** Simulated current-space map obtained from the TDGL simulations. **f** Snapshots of the order parameter when the near-field hotspot is placed close to the edge (situation I) and when the near-field hotspot is placed on top of the defect (situation II). Arrows indicate the direction of movement of the vortices.

Having established that the local photovoltage peaks arise from defects, we now turn to the response at the edges, which provides insight into the underlying $V_{\mathrm{PV}}$ generation mechanism. Fig. 4a shows current-space maps taken at three different temperatures. In all cases, similar to what was shown in Fig. 2 and Fig. 3, we observe an increase in $V_{\mathrm{PV}}$ towards the edge of the strip. The evolution of $V_{\mathrm{PV}}$ at the edge with current (Fig. 4b) shows a strong non-monotonic dependence, similar to that observed for the response of defects. To further characterize this behavior, we compare these traces with the evolution of the differential resistance, $\mathrm{d}V/\mathrm{d}I$, obtained from DC transport under far-field THz illumination taken when the tip is retracted (Fig. 4c). The edge $V_{\mathrm{PV}}$ signal onsets close to the critical current ($I_c$) determined from DC transport (short-dashed lines in Fig. 4a). Above $I_c$, the edge signal increases, reaching a maximum at a current slightly below the maximum of the $\mathrm{d}V/\mathrm{d}I$ (long-dashed lines in Fig. 4b). Increasing the bias current further reduces the $V_{\mathrm{PV}}$ signal, as the strip transitions into the normal state, characterized by a resistance $R_{\mathrm{N}}$. Both the $V_{\mathrm{PV}}$ traces and the differential resistance trends broaden in current as temperature is increased.

The close correspondence between the trend observed in $V_{\mathrm{PV}}$ and the differential resistance suggests a common physical origin. In type-II superconductors such as NbN, a resistive state appears above the critical current due to the onset of vortex motion[23–25]. In the absence of magnetic field, current-induced vortices and antivortices form at opposing edges of the strip and traverse the sample perpendicular to the current direction. Their motion produces time-dependent phase slips that, via the Josephson relation, generate a voltage along the current direction[26]. Since vortex dynamics govern this resistive state, any local perturbation that modifies vortex nucleation or motion should produce a measurable voltage change. The THz near-field provides precisely such a perturbation: the tip confines THz radiation into a nanoscale volume, where photons with energy above the superconducting gap break Cooper pairs and suppress the local order parameter –creating, in effect, a nanoscale hotspot. We therefore attribute the measured $V_{\mathrm{PV}}$ to the interaction between this near-field induced hotspot and current-induced vortices. Importantly, the interaction between vortices and light is known to play a significant role in the response of superconducting photodetectors[27,28].

We confirm this interpretation performing time-dependent Ginzburg-Landau (TDGL) simulations of a current-carrying superconducting strip close to $T_c$. The simulated strip contains a defect close to the center and a hotspot whose position is varied across the width of the strip (more details in *Methods*). The simulations were performed at $T/T_c = 0.98$. To model the near-field induced hotspot we introduced a circular region with radius $r_{hotspot} = 196$ nm and a local temperature 1% above the bath temperature. The defect was included as a circular region with radius $r_{defect} = 118$ nm and local $T_c$ 1% lower than the bulk $T_c$. We compare our TDGL simulation results to the experimental current-space map taken at $T = 13.86$ K containing one local photovoltage peak close to the center (Fig. 4d). A constant background, taken as the average of the signal outside of the strip, was subtracted from the experimental map to facilitate the comparison. The simulated current-space map (Fig. 4e) reproduces the main features observed in the experimental current-space maps: an increased signal at the edges whose magnitude decreases towards the bulk, and a local photovoltage peak at the position of the defect.

These features can be explained by examining the time evolution of the order parameter and the effects arising from its suppression by the near-field hotspot. This local suppression by the hotspot forces the supercurrent to bend around it, leading to local current crowding and thereby enhancing the force acting on the vortices. Additionally, when the hotspot overlaps with the edge (Fig. 4f, Situation I) it lowers the vortex entry barrier, thereby allowing vortices to penetrate at lower bias currents. Together, these two effects promote vortex penetration, resulting in an enhanced signal at the edges.

The same current crowding mechanism accounts for $V_{\mathrm{PV}}$ generation when the hotspot does not overlap with the edge. In this case, the induced curl in the supercurrent generates a magnetic dipole at the hotspot position, leading to the formation of vortex-antivortex pairs (V-AV pairs)[27,29,30]. These pairs are launched once the current density and supervelocity in the hotspot become sufficient to unbind them, producing a measurable $V_{\mathrm{PV}}$ signal. As the hotspot moves away from the edge, current crowding weakens, explaining the decay in $V_{\mathrm{PV}}$ towards the bulk. This phenomenology is in agreement with reports of spatially resolved responsivity in superconducting photodetectors, as well as theoretical predictions from vortex-assisted detection models[2,27,28,31–33].

When the hotspot overlaps with a defect (Fig. 4f, Situation II), V-AV pair creation is enhanced. This is due to the additional suppression of the order parameter provided by the defect, which further reduces the V-AV binding energy. Overall, the creation of V-AV pairs inside the near-field-induced hotspot is controlled by three different factors: the injection current, the distance from the edge, and the local superfluid density.

The current dependence of $V_{\mathrm{PV}}$ can be attributed to the evolution of the vortex state in the strip. Close to the critical current ($I_c$), vortices at the edge are highly sensitive to perturbations that reduce the entry barrier, explaining the onset of the photovoltage near this bias. Once the current has surpassed this value the strip enters the flux-flow regime, where the voltage drop across the strip is determined by the vortex velocity. In this case, incrementing the injection current leads to an enhancement of V-AV pair creation and an increase in their velocity, which translates to an increase of $V_{\mathrm{PV}}$, as shown in the simulated current-space map (Fig. 4e).

In the simulated map we do not reach the downturn in $V_{\mathrm{PV}}$ observed in the experiment. One possible reason is that our simulations are constrained to current values below the flux-flow instability. This point, signalled by the maximum of the $\mathrm{d}V/\mathrm{d}I$, is characterized by the appearance of quasi-phase-slip lines that appear when the moving vortices start merging[25,34,35]. As shown in Fig. 4c, the downturn in $V_{\mathrm{PV}}$ starts close to this point, suggesting that the near-field photovoltage decreases when the strip crosses the flux-flow instability. Given that beyond this point dissipation becomes dominated by the quasi-phase-slip-lines dynamics rather than by vortex launching, the influence of the tip-induced hotspot diminishes, explaining the downturn of $V_{\mathrm{PV}}$. Upon reaching the normal state, vortices are no longer present and $V_{\mathrm{PV}}$ vanishes. This is clearly observed at $T = 13.82$ K, where the signal suddenly drops after the strip sharply transitions to the normal state.

# Conclusion

In this study, we have demonstrated that THz near-field photovoltage measurements are capable of spatially mapping inhomogeneities present in a superconductor and measure their vortex-assisted photoresponse with nanometric resolution. The average size of the defect analyzed in Fig. 2, $\langle \Delta \mathrm{x} \rangle = 300 \pm 30$ nm, is comparable to the tip radius used in the experiment, suggesting that our resolution is limited by tip size, as expected for near-field measurements. This resolution could be further improved using sharper tips, potentially pushing the resolution limit below 100 nm and enabling further valuable insights into light-matter interaction in superconducting materials.

Compared to low-temperature laser scanning microscopy (LTLSM), which is typically limited to resolutions above 1 µm[36,37], our technique improves the spatial resolution by at least a factor of 3. The crucial difference between THz near-field photovoltage and LTLSM lies in the photon energy used: the energy of 2.52 THz radiation (10.4 meV) is more than two orders of magnitude lower than that of visible photons employed in LTLSM, making THz near-field photovoltage measurements less disruptive to the superconducting state. As a consequence, we are able to explore the vortex-assisted photoresponse of defects, revealing a strong non-monotonic dependence on current-bias that correlates with the evolution of the vortex-dissipative state in the film. Notably, the observed maximum in $V_{\mathrm{PV}}$ close to the vortex flow instability indicates that THz near-field photovoltage is sensitive to the transition between different vortex regimes, offering a local probe complementary to the study of global transport.

Within the proposed detection mechanism, an enhanced response is expected at defects where superconductivity is locally suppressed, consistent with the behavior reported in Fig. 2 and Fig. 3. While a direct, one-to-one correlation between near-field maps and electron microscopy is not possible, complementary SEM and energy-dispersive X-ray spectroscopy measurements reveal several Nb-depleted regions with lateral dimensions ranging from 150 to 700 nm (Supplementary Note 2). These length scales are comparable to the spatial extent of the observed localized $V_{\mathrm{PV}}$ peaks, making such regions plausible candidates for sites of enhanced V-AV pair generation. An additional contribution to defects may arise from the intrinsic granularity of the film, which exhibits grain sizes of order 30 nm and comparable intergrain spacing (Supplementary Note 1). In this case, the observed size of the photovoltage features would be limited by the resolution of the measurement.

THz near-field photovoltage offers an alternative route to probing the superconducting state locally, giving access to its local electromagnetic response. Other scanning probes, such as magnetic force microscopy, scanning NV magnetometry, and scanning SQUID, can detect static magnetic flux from vortices with comparable spatial resolution, but have limited sensitivity to dynamic, light-driven responses. Scanning tunnelling microscopy can map the superconducting gap with atomic resolution in equilibrium, but is blind to dynamical responses at THz frequencies and is not easily operated under finite in-plane bias. What distinguishes THz near-field photovoltage is the combination of all three: local access, excitation at the natural energy scale of the condensate, and compatibility with non-equilibrium transport regimes. More broadly, its sensitivity to local variations in superfluid density opens new opportunities to study light–matter interactions at low energies in inhomogeneous superconductors, particularly in high-Tc materials and moiré systems, where the nanoscale coexistence of competing correlated phases lies at the heart of the physics.

# Methods

## Sample fabrication and microstructure characterization

The NbN film with thickness 320 nm was sputtered at 250 °C in a $N_2$ + Ar atmosphere, at 5 mTorr pressure, onto a high-resistivity silicon wafer with 370 nm thermal $SiO_2$. After sputtering, the geometry was defined using photolithography and reactive-ion etching with the mixture of $CHF_3$ and $SF_6$ gases. Lastly, the wafer was spin-coated with photoresist before dicing and shipment. Once received, the samples were sonicated in acetone and isopropanol to remove the photoresist and expose the clean surface. SEM imaging was performed with ZEISS Gemini 560 in the secondary electron scattering mode.

## THz Near-field photovoltage nanoscopy

2.52 THz continuous-wave light was generated using a commercial THz gas laser (Edinburgh Instruments FIRL-100). The THz light was coupled to a commercial cryoSNOM system (Neaspec cryo-neaSCOPE). We used Au-coated AFM tips (Team Nanotec LRCH) with large tip radius (~ 250 nm) and resonance frequency $\Omega \approx 60$ kHz. The Neaspec AFM controller incorporates a damping system that allows for the lowering of the quality factor of the tip resonance. This ensures stable AFM tapping mode at cryogenic temperatures. DC current was injected using a source meter (Keithley 2470). The near-field photovoltage was first filtered (bandpass 3-100 kHz) and enhanced with an Ithaco pre-amplifier and then recorded using the digital lock-in amplifier available in the Neaspec cryo-neaSCOPE interface.

## Electron transport characterization

B-field electron transport was measured in an ICE Oxford cryostat with standard lock-in and DC measurement techniques. All the IV curves were collected in the current-bias mode. Zero-field IV curves in the dark and under illumination were taken in the cryoSNOM before each near-field scan.

## Simulation details

The simulations were carried out through the numerical solution of the two-dimensional generalized time-dependent Ginzburg-Landau equation (TDGL)[38]:

$$\frac{u}{\sqrt{1+\gamma^2|\psi|^2}}\left(\frac{\partial}{\partial t}+i\varphi+\frac{\gamma^2}{2}\frac{\partial|\psi|^2}{\partial t}\right)\psi=(\nabla-i\boldsymbol{A})^2\psi+\psi\left(1-\frac{T(\boldsymbol{r})}{T_c(\boldsymbol{r})}-|\psi|^2\right).$$

The order parameter $\psi$ is in units of the field-free order parameter $\psi_\infty$ at $T = 0$; lengths in units of the superconducting coherence length $\xi(0)$; time in units of $t_{GL} = \pi\hbar/8uk_bT_c$, where $u = 5.79$ comes from microscopic theory; the parameter $\gamma$ is proportional to the product inelastic electron-collision time $\tau_e$ and $\psi_\infty$, being given by $\gamma = 2\tau_e\psi_\infty/\hbar$. Here, we have fixed $\gamma = 10$. The vector potential $\boldsymbol{A}$ in units of $H_{c2}(0)\xi(0)$, with $H_{c2}(0)$ being the upper critical field at $T = 0$; the current density is in units of $J(0) = \sigma_n\varphi(0)/\xi(0)$, where $\sigma_n$ is the material normal conductivity; and the electrostatic potential $\varphi$ in units of $\varphi(0) = \hbar/2et_{GL}$. At each time step, $\varphi$ is obtained from the Poisson equation

$$\nabla^2\varphi = \nabla\cdot\boldsymbol{J_s},$$

Where the supercurrent is given by:

$$\boldsymbol{J_s} = \mathrm{Im}[\bar{\psi}(\nabla - i\boldsymbol{A})\psi].$$

Boundary conditions for $\psi$ are set to guarantee no supercurrent flows out of the film. The bias current density $\boldsymbol{J_b}$ is sourced along the y direction through the boundary condition for $\varphi$, with $\nabla\varphi = j_\mathrm{b}$ along the surfaces perpendicular to the external current, while $\nabla\varphi = 0$ is set at the remaining surfaces. The simulated strip had dimensions $w = 1120\xi(0)$ and $L = 890\xi(0)$. For practical reasons, the ratio $w/\xi(0)$ is limited to $\sim 10^3$, whereas in our experiment it is closer to $10^4$. Achieving the experimental $w/\xi(0)$ ratio would lead to excessively long computation times, as the TDGL equations are solved on a finite mesh with cell area $\sim \xi(0)^2$. Details on the numerical discretization can be found in Refs [35,39].

## Acknowledgements

The authors are thankful to D. Yu. Vodolazov, A. G. Sivakov, M. van Exter and C. Feuillet-Palma for the discussions on vortex physics, and I. Iorsh and L. Bishop-Van Horn for the discussions on the vortex dynamics simulations. E.K. acknowledges Marie-Sklodowska-Curie Fellowship, Project SUPERTERA, Ref. 101033521. F.H.L.K. acknowledges support from the Gordon and Betty Moore Foundation through Grant GBMF12212, and the Government of Spain (QTWIST RD0768/2022, PID2022-141081NB-I00; PRE2022-101705, Severo Ochoa CEX2024- 001490-S, MCIN/AEI/10.13039/501100011033). This work was also supported by the European Union NextGenerationEU / PRTR(PRTR-C17.I1), PCI2021-122020-2A within the FLAG-ERA grant [PhotoTBG], and EXQIRAL 101131579, Fundació Cellex, Fundació Mir-Puig, Generalitat de Catalunya (CERCA, Department of Digital Policies and Territory, AGAURAn Electronic Quantum Simulator – BBVA Foundation 2022 Fundamentals Program. Views and opinions expressed are those of the author(s) only and do not necessarily reflect those of the European Union Research Executive Agency. Neither the European Union nor the granting authority can be held responsible for them. This material is based upon work supported by the Air Force Office of Scientific Research under Award Number FA8655-23-1-7047. Any opinions, findings, conclusions, or recommendations expressed in this material are those of the author(s) and do not necessarily reflect the views of the United States Air Force. Work performed at the Center for Nanoscale Materials, a U.S. Department of Energy Office of Science User Facility, was supported by the U.S. DOE, Office of Basic Energy Sciences, under Contract No. DE-AC02-06CH11357. L.R.C. and M.V.M. acknowledge funding from the Research Foundation-Flanders (FWO Vlaanderen) and the US-Army Research Office under Grant Number W911NF-24-1-0145.